\documentclass{aa}

\usepackage{graphicx}
\usepackage[varg]{txfonts}
\usepackage{color}

\begin{document}

\title{A key factor to the spin parameter of uniformly rotating compact stars: crust structure}

\author{B. Qi\inst{1,2}
\and N. B. Zhang\inst{1}
\and B. Y. Sun\inst{3,4}
\and S. Y. Wang\inst{1}
\and J. H. Gao\inst{1}}

\institute{Shandong Provincial Key Laboratory of Optical Astronomy and
Solar-Terrestrial Environment, School of Space Science and Physics, Shandong University, Weihai 264209, China
\and Department of Physics, University of Notre Dame, Notre Dame, Indiana 46556, USA
\and School of Nuclear Science and Technology, Lanzhou University, Lanzhou 730000, China
\and Key Laboratory of Special Function Materials and Structure Design, Ministry of Education, China\\
\email{sunby@lzu.edu.cn}}



\abstract
{}
{We study the key factor to determine the  dimensionless spin parameter $j\equiv cJ/(GM^2)$ of different kinds of uniformly
rotating compact stars, including the traditional neutron stars, hyperonic neutron stars, and hybrid stars,
and check the reliability of the results on various types of equations of state of dense matter.}
{The equations of state from the relativistic mean field theory and the MIT bag model are adopted to simulate compact stars.
Numerical calculations of rigidly rotating neutron stars are performed using the RNS code in the framework of general relativity  by solving the Einstein equations for stationary axis-symmetric spacetime. }
{The crust structure of compact stars is found to be a key factor to determine the maximum value of the spin parameter
$j_{\rm max}$. For the stars with inclusion of the crust, $j_{\rm max}\sim 0.7$ is sustained for various kinds of compact stars
with $M>0.5 M_{\odot}$, and is found to be insensitive to the mass of star and selected equations of state. For the traditional
neutron stars, hyperonic neutron stars and hybrid stars without crust structure, the value $j_{\rm max}$ lies in the range of $[0.7, 1.0]$.
Thus, not $j>0.7$ but $j>1$ could be treated as the candidate criterion to distinguish the strange quark stars from the other kinds of compact stars.
Furthermore, a universal formula $j=0.48(f/f_K)^3-0.42(f/f_K)^2+0.63(f/f_K)$ is suggested to calculate the spin parameter
at any rotational frequency for all kinds of compact stars with crust structure and $M>0.5M_{\odot}$.}
{}

\maketitle

\section{Introduction}

Compact stars, as one of the most exotic objects in the universe, play the role of a bridge among astrophysics, nuclear physics and particle physics. The interior of the compact stars is still a mission to understand, which may contain a mixture of exotic particles based on the physics of strong interactions such as the strangeness-bearing baryons \citep{Glendenning85,Glendenning97}, condensed mesons \citep{Rho79, Kaplan86, Kaplan86b}, or even deconfined quarks \citep{Collins75}. In theoretical studies, the compact stars could be divided generally into several types by assuming the different components inside: traditional neutron stars (composed by $\beta$-equilibrium nucleon matter), hyperonic neutron stars (including strangeness-bearing baryons), hybrid stars (with hadron-quark phase transition) and strange quark stars (composed by up, down and strange quarks) and so on, as shown in Fig.~1 of \citet{Weber07}. How to identify the type and the inner structure of existent compact stars via the observable quantities is always a great challenge.


In the past decades, many millisecond pulsars have been reported~\citep{Backer82,2,3}. Various numerical codes have been developed recently to construct the models for rapidly rotating compact stars in general relativity~\citep{4}. As directly measurable quantities, the rotational frequency $f$ of pulsars and its possible maximum value, namely the keplerian frequency $f_K$, have attracted most of interests in previous studies of rotating compact stars~\citep{5,6,7}. Recently, another important characteristic quantity, i.e., the dimensionless spin parameter $j\equiv cJ/(GM^2)$, was introduced into the investigative territory of rotating compact stars~\citep{8}, where $J$ is the angular momentum and $M$ is the gravitational mass of the stars. It is suggested that the spin parameter $j$ could play an important role in understanding the observed quasi-periodic oscillations (QPOs) in disk-accreting compact-star systems~\citep{Miller98,10} and the final fate of the collapse of a rotating compact star~\citep{8}. In \citet{8}, it was revealed that the maximum value of the spin parameter $j_{\rm max}$ of a neutron star rotating at the keplerian frequency has a upper bound of about 0.7, which is essentially independent on the mass of neutron star as long as the mass is larger than about $1M_{\odot}$. However, the spin parameter of a quark star described by the MIT bag model does not have such a universal upper bound and could be larger than unity. Thus, the spin parameter extracted from the observable could provide a strict and new constraint on the inner structure of compact stars and corresponding equation of state (EOS) of dense matter. To utilize this constraint in more future works, its physical origin, e.g., the reason why $j_{\rm max}\thicksim0.7$ for rotating neutron stars, and the reliability on various types of EOSs of dense matter need further study.

In this paper, the dimensionless spin parameter $j$ of three types of rotating compact stars, namely, the traditional neutron stars, hyperonic neutron stars and hybrid stars, will be studied using the equations of state from the relativistic mean field (RMF) theory~\citep{11,12,13} and the MIT bag model~\citep{Chodos74, Fahri84, Witten84}. The essential factor to determine the maximum value of the spin parameter $j_{\rm max}$ will be investigated. The paper is organized as follows: In Section 2 we briefly introduce the numerical method and the EOS models used in the calculations. The bulk and characteristic properties of different kinds of compact stars without or with rotation will be analyzed in Section 3. Section 4 will show our results on the key factor to determine the $j_{max}$ and the corresponding astrophysical implications. Finally a short summary will be given.

\section{Numerical Method and EOS Models}
We compute numerically the uniformly rotating compact stars with the RNS code (http://www.gravity.phys.uwm.edu/rns/), which solves the hydrostatic and Einstein's field equations for the rotating rigidly stars under the assumption of stationary and axial symmetry about the rotational axis, and reflection symmetry about the equatorial plane. For the details of the model we refer the reader to \citet{4,5,25,26} and references therein. The angular momentum $J$ of a rotating compact star is calculated in the code as follows
\begin{equation}
    J=\int T^{\mu\nu}\xi^{\nu}_{(\phi)}dV~,
\label{J}
\end{equation}
where $T^{\mu\nu}$ is the energy-momentum tensor of stellar matter, $\xi^{\nu}_{(\phi)}$ is the Killing vector in the azimuthal direction reflecting axial symmetry, and $dV$ is a proper 3-volume element.

For the neutron stars (NS) with hadronic matter, we adopt two typical effective interactions of the density-dependent RMF theory, i.e., TW99~\citep{27} and PKDD~\citep{28} to model the EOSs without or with hyperons (denoted as TW99N and PKDDN for traditional NS and TW99H and PKDDH for hyperonic NS respectively). As illustrated in~\cite{Hofmann01,Ban04,Sun08,Long12,Zhang13}, a self-consistent and successful description on the properties of nuclear matter at a large range of density and the observable of neutron stars could be obtained via a microscopic effective Lagrangian founded on the meson exchange diagram. For the hyperonic neutron stars, the hyperons $\Lambda$, $\Sigma^{\pm}$, $\Sigma^0$, $\Xi^-$, $\Xi^0$ are covered beyond the traditional NS, and the meson-baryon coupling constants are chosen according to~\cite{Ban04}. For the hybrid star models, we adopt the MIT bag models~\citep{Chodos74, Fahri84, Witten84} with the bag constant $B=90, 150$ MeV fm$^{-3}$ for the quark matter and PKDDN from RMF for the hadronic matter, which are denoted as Hybrid90, Hybrid150 respectively~\citep{EXOCT07}. In the MIT bag model, we include massless $u$ and $d$ quarks as well as the $s$ quark with the mass $m_s = 150$ MeV. The hadron-quark phase transition of the hybrid stars is considered by the Gibbs (Glendenning) construction~\citep{Glendenning92,Nicotra06}.

It is easy to imagine that the matter distribution inside a star is essential to the momentum of inertia while the star mass is fixed. Thus, the crust structure of a star should play an important role in its angular momentum $J$. In general, the crust of neutron stars can be separated into two parts: outer crust and inner crust~\citep{Lattimer04}. The main compositions of the outer crust are ions and electrons. The density of the bottom of this part is about $\rho=4 \times 10^{11} {\rm g~cm^{-3}}$. On the other hand, the neutrons drip out of the nuclei in the inner crust which consists of electrons, free neutrons and neutron-rich nuclei. The density of the interface between inner crust and core is about $0.5~\rho_0$ (i.e., $0.08~\rm{fm}^{-3}$), where $\rho_0$ is the so called saturation density of nuclear matter. To consider the effects of the crust, two sets of crust EOS are chosen in the low-density region ($\rho < 0.08~\rm{fm}^{-3}$) instead of RMF or MIT calculations: Set 1, the Negele and Vautherin (NV) EOS \citep{Negele73} for the inner crust and Baym-Pethick-Sutherland (BPS) EOS \citep{Baym71} for the outer crust; Set 2, the Haensel-Pichon(HP) EOS \citep{HP94} for the inner crust and Douchin-Haensel (DH) EOS \citep{Douchin01} for the outer crust.

\begin{figure}[h]
\centering
\includegraphics[width=0.45\textwidth]{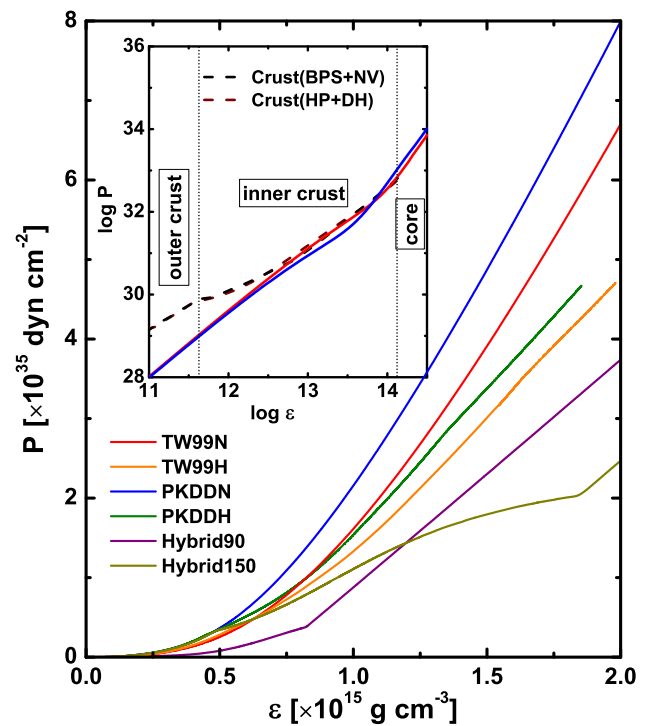}
\caption{The equations of state for the compact star matter adopted in the present work,
namely the pressure as a function of the energy density for the traditional neutron stars (PKDDN and TW99N), hyperonic neutron stars (PKDDH and TW99H) and hybrid stars (Hybrid90 and Hybrid150), respectively. Plot inserted: the adopted equations of state for the crust matter (BPS+NV and HP+DH) at low densities, including the outer crust and the inner crust, which are exhibited in the logarithmic coordinate in comparison with PKDDN and TW99N. The dotted lines in the inserted plot show the boundaries between inner crust and outer crust, inner crust and core, respectively.}
\label{fig:1}
\end{figure}

In Fig.~\ref{fig:1}, we present the EOSs for different kinds of compact stars, namely, PKDDN and TW99N for the traditional NS, PKDDH and TW99H for the hyperonic NS, as well as Hybrid90 and Hybrid150 for the hybrid stars, respectively. Two crust EOSs, namely, NV+BPS and HP+DP are exhibited in the inserted plot in comparison with PKDDN and TW99N, where a logarithmic coordinate is used to show the difference more clearly. The dotted lines in the inserted plot show the boundaries between inner crust and outer crust, inner crust and core, respectively. It is illustrated that both two crust EOSs have a stiffer behavior at very low densities than PKDDN and TW99N composed by $\beta$-equilibrium nucleon matter. Therefore, the inclusion of the crust should be significant in understanding the physics related to the surface of compact stars.

\begin{table*}
\caption{The gravitational mass $M^{\rm sta}_{\rm max}$, the equatorial radius $R^{\rm sta}_{\rm max}$ and the central density $\varepsilon_c^{\rm sta}$ for the non-rotating compact stars with maximum mass configuration, and the corresponding values  $M^{\rm kep}_{\rm max}$, $R^{\rm kep}_{\rm max}$ and $\varepsilon_c^{\rm kep}$ for the keplerian rotating compact stars with maximum mass configuration. The keplerian frequency $f_{K}$ and corresponding spin parameter $j_{\rm max}$ are also given. The values  out of/in the brackets denote the results with/without crust structure.}
\label{Tab1}
\centering
\begin{tabular}{c|ccc|ccccc}
    \hline\hline
     \noalign{\smallskip}
       EOS     & $M^{\rm sta}_{\rm max}$ & $R^{\rm sta}_{\rm max}$ & $\varepsilon_c^{\rm sta}$ & $M^{\rm kep}_{\rm max}$ &$R^{\rm kep}_{\rm max}$&$\varepsilon_c^{\rm kep}$&$f_{K}$      & $j_{\rm max}$\\
               &      $(M_{\odot})$      &    (${\rm km}$)         & ($10^{15}{\rm g/cm^3}$)   &        $(M_{\odot})$    &      (${\rm km}$)     &($10^{15}{\rm g/cm^3}$)  & (${\rm kHz}$)&    \\
     \noalign{\smallskip}
    \hline
     \noalign{\smallskip}
    TW99N    & 2.08~(2.08) & 10.62~(10.63) & 2.58~(2.38) & 2.48~(2.52) & 14.16~(14.24) & 2.24~(2.09) & 1.67~(1.68) & 0.67~(0.70)\\
    PKDDN    & 2.33~(2.34) & 11.78~(11.73) & 2.08~(1.98) & 2.78~(2.84) & 15.69~(15.70) & 1.80~(1.72) & 1.51~(1.53) & 0.67~(0.70)\\
    TW99H    & 1.83~(1.82) & 10.75~(10.97) & 2.50~(2.29) & 2.18~(2.23) & 14.60~(14.78) & 2.19~(2.36) & 1.51~(1.51) & 0.64~(0.68)\\
    PKDDH    & 1.97~(1.98) & 11.48~(11.33) & 2.28~(2.35) & 2.34~(2.40) & 15.54~(15.53) & 1.97~(1.85) & 1.42~(1.45) & 0.64~(0.68)\\
    Hybrid90 & 1.51~(1.52) &~~8.99~(~~8.97)& 3.41~(3.26) & 1.84~(1.86) & 12.36~(12.29) & 2.79~(2.84) & 1.78~(1.82) & 0.67~(0.69)\\
    Hybrid150& 1.50~(1.50) & 12.29~(11.93) & 1.85~(1.79) & 1.81~(1.88) & 17.13~(16.94) & 1.61~(1.46) & 1.10~(1.15) & 0.63~(0.70)\\
     \noalign{\smallskip}
    \hline\hline
\end{tabular}
\end{table*}

\begin{table*}
\caption{The equatorial radius $R^{\rm kep}$, the central density $\varepsilon_c^{\rm kep}$, the keplerian frequency $f_{K}$ and corresponding spin parameter $j_{\rm max}$ for the keplerian rotating compact stars with the bayonic mass $M_{b}=1.0$ $M_{\odot}$ (left half) and 1.4 $M_{\odot}$ (right half).  The values  out of/in the brackets denote the results with/without crust structure. }
\label{Tab2}
\centering
\begin{tabular}{c|cccc|ccccc}
    \hline\hline
     \noalign{\smallskip}
       EOS     &$R^{\rm kep}_{1.0}$&$\varepsilon^{\rm kep}_{c}$ &$f_{K}$ & $j_{\rm max}$ &$R^{\rm kep}_{1.4}$&$\varepsilon^{\rm kep}_{c}$ &$f_{K}$ & $j_{\rm max}$ \\
            &(${\rm km}$)           &($10^{15}{\rm g/cm^3}$)         & (${\rm kHz}$)  &  & (${\rm km}$)           &($10^{15}{\rm g/cm^3}$)         & (${\rm kHz}$)  & \\
     \noalign{\smallskip}
    \hline
     \noalign{\smallskip}
    TW99N     & 17.90~(17.10) & 0.54~(0.51) & 0.75~(0.83) & 0.68~(0.85) & 17.64~(17.22) & 0.66~(0.62) & 0.90~(0.95) & 0.68~(0.80)\\
    PKDDN     & 19.95~(18.92) & 0.41~(0.38) & 0.64~(0.71) & 0.69~(0.91) & 19.68~(19.15) & 0.50~(0.47) & 0.77~(0.82) & 0.69~(0.84)\\
    TW99H     & 18.09~(17.35) & 0.52~(0.41) & 0.74~(0.81) & 0.69~(0.86) & 17.78~(17.60) & 0.65~(0.59) & 0.88~(0.92) & 0.69~(0.81)\\
    PKDDH     & 19.77~(18.82) & 0.41~(0.37) & 0.65~(0.72) & 0.70~(0.92) & 19.58~(18.90) & 0.51~(0.46) & 0.77~(0.82) & 0.70~(0.85)\\
    Hybrid90  & 13.43~(13.61) & 0.97~(0.95) & 1.12~(1.14) & 0.67~(0.74) & 13.59~(13.65) & 1.14~(1.12) & 1.30~(1.32) & 0.69~(0.73)\\
    Hybrid150 & 19.66~(18.83) & 0.42~(0.39) & 0.66~(0.72) & 0.70~(0.90) & 19.43~(19.04) & 0.54~(0.48) & 0.78~(0.83) & 0.70~(0.84)\\
     \noalign{\smallskip}
    \hline\hline
\end{tabular}
\end{table*}

\section{Properties of Different Kinds of Compact Stars}

The bulk properties of compact stars adopting the above EOSs are calculated and presented in Tab.~\ref{Tab1}, including the gravitational mass $M^{\rm sta}_{\rm max}$, the equatorial radius $R^{\rm sta}_{\rm max}$ and the central density $\varepsilon_c^{\rm sta}$ for the non-rotating stars with maximum allowable mass, and the corresponding values for the keplerian rotating stars with the maximum allowable mass. The keplerian frequency $f_K$ and corresponding spin parameter $j_{\rm max}$ are also given. The values out of/in the brackets denote the results with/without crust structure. It has been checked that two sets of crust models almost give the same results, thus only those with NV+BPS crust structure are listed here. For comparison, Tab.~\ref{Tab2} shows the results for the keplerian rotating compact stars with the fixed bayonic mass $M_{b}=1.0$ $M_{\odot}$  and 1.4 $M_{\odot}$. One could see these bulk properties depend on the chosen EOSs sensitively, in both static and keplerian rotating cases. Taking $M^{\rm sta}_{\rm max}$ as an example, the values alter from 1.5$M_{\odot}$ to about 2.3$M_{\odot}$. On the contrary, it is revealed that the influence of the crust on the observable quantities of compact stars is quite small, especially for the cases of maximum allowable mass (see Tab.~\ref{Tab1}) where the maximum deviation is found to be $1\%$ for $M^{\rm sta}_{\rm max}$, 3\% for $R^{\rm sta}_{\rm max}$, 4\% for $M^{\rm kep}_{\rm max}$, 1\% for $R^{\rm kep}_{\rm max}$ and 5\% for $f_K$. The deviation becomes a little larger as the bayonic mass of stars goes smaller, but still keep a small value, e.g., for the compact stars with $M_{b}=1.0$ $M_{\odot}$ about 5\% for $R^{\rm kep}_{\rm 1.0}$ and 10\% for $f_K$ (see Tab.~\ref{Tab2}). As an exception, the difference in the maximum value of the spin parameter $j_{\rm max}$ due to the crust effects becomes non-negligible when the bayonic mass of stars decreases, namely, 18\%~(24\%) for the stars with $M_{b}=1.4~(1.0)$ $M_{\odot}$. Thus, the sensitivity of $j_{\rm max}$ to the crust structure makes it to be a possible well characteristic quantity in understanding the physics related to the surface of compact stars.


\begin{figure}[h]
\centering
\includegraphics[width=0.48\textwidth,bb=0 0 680 410]{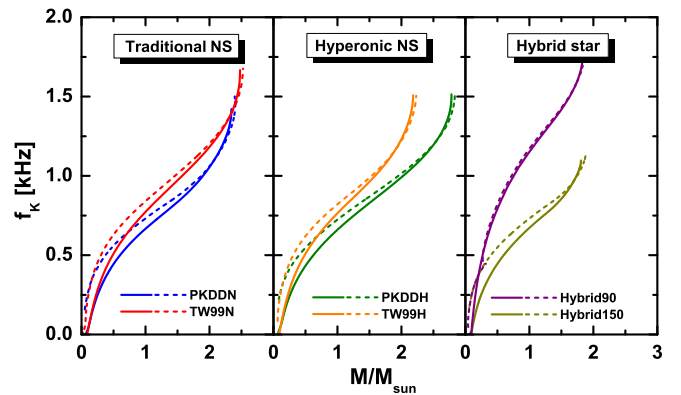}
\caption{The keplerian frequency $f_K$  as a function of the gravitational mass for the traditional neutron stars (PKDDN and TW99N), hyperonic neutron stars (PKDDH and TW99H) and hybrid stars (Hybrid90 and Hybrid150), respectively. The solid (dashed) lines represent the results with (without) crust structure.}
\label{fig:2}
\end{figure}

As fundamental quantity to describe the rapidly rotating compact stars, the keplerian frequency $f_K$ is presented in Fig.~\ref{fig:2} at various gravitational masses for the traditional neutron stars (PKDDN and TW99N), hyperonic neutron stars (PKDDH and TW99H) and hybrid stars (Hybrid90 and Hybrid150), respectively. The solid (dashed) lines represent the results with (without) crust structure. It can be found the influence of crust structure on the relationship of $f_K\sim M$ is not large, in agreement with the above discussions. To be more precise,  there are an approximate empirical equation $f_K(M)= C(M/M_\odot)^{1/2}(R_S/10~{\rm km})^{-3/2}$  among $f_K$, the gravitational mass $M$ and the radius $R_S$ of the non-rotating static star with the same mass $M$ \citep{Lattimer04,7}.  \citet{7} suggested $C= 1.08~$kHz  for the neutron stars and $C=1.15~$ kHz for the quark star, when $0.5M_{\odot}<M<0.9M^{\rm stat}_{\rm
max}$. In \citet{Zhang13}, we have systematically checked the empirical relationship based on the various RMF EOSs without and with hyperon, and found it is a good approximation in all of the cases. In the present work, such an empirical relationship with $C=1.08~$kHz is also proved to be applicable to the hybrid stars, which is not discussed before. In fact, it is found the relationship $f_K(M)= 1.08 {\rm kHz} (M/M_\odot)^{1/2}(R_S/10~{\rm km})^{-3/2}$ are good approximation  for both cases in Fig.~\ref{fig:2} without and with the crust structure, and the deviation of the values from the equation is within about $2\%$ for $0.5M_{\odot}<M<0.9M^{\rm stat}_{\rm
max}$.

\section{The Crust Effects on Dimensionless Spin Parameter}

Now we turn to the dimensionless spin parameter $j\equiv cJ/(GM^2)$. \citet{8} has shown that $j_{max} \sim0.7$ for neutron stars without hyperons, while the spin parameter of a quark star modeled by the MIT bag model can be larger than unity and does not have a universal upper bound. It is interesting to further explore what the key factor to determine the spin parameter is, and study the reliability of the result on various selected EOSs of dense matter, such as hyperonic NS and hybrid stars, which are not considered yet in \citet{8}. The calculations and the results for strange quark stars with the MIT bag model are very similar to those in \citet{8}, which have $j_{max}>1$ for most of star masses, thus we do not exhibit here.

\begin{figure}
\centering
\includegraphics[width=0.48\textwidth]{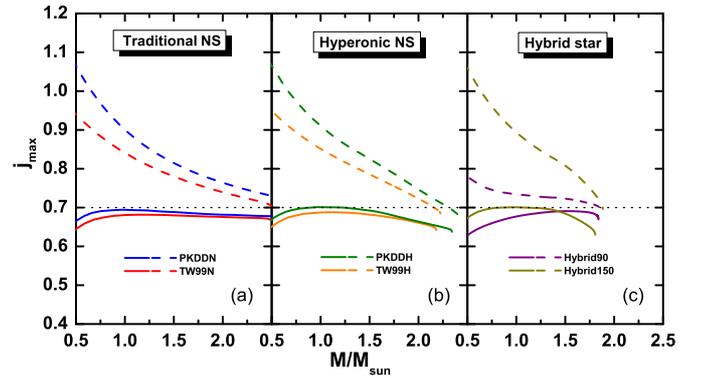}
\caption{
The maximum value of the dimensionless spin parameter $j_{\rm max}$ of a rotating compact star as a function of the gravitational mass  for the traditional neutron stars (PKDDN and TW99N), hyperonic neutron stars (PKDDH and TW99H) and hybrid stars (Hybrid90 and Hybrid150). The solid (dashed) lines represent the results with (without) crust structure. The reference value $j_{\rm max}=0.7$ is shown as well with the dotted line.}
\label{fig:3}
\end{figure}

The maximum value of the spin parameter $j_{max}$ of a rotating compact star for the traditional neutron stars (PKDDN and TW99N), hyperonic neutron stars (PKDDH and TW99H) and hybrid stars (Hybrid90 and Hybrid150) is shown in Fig.~\ref{fig:3} as a function of the gravitational mass. The solid (dashed) lines represent the results with (without) crust structure. Here we adopt the NV+BPS as the crust EOS. As shown in Fig.~\ref{fig:3}(a), when the crust is included, $j_{max}\sim 0.7$ for the traditional NS as long as $M>0.5M_{\odot}$, while the behavior of $j_{max}$ is quite different if there is no crust. For the neutron star without the crust structure, which means that the neutron star is made of $\beta$-equilibrium nucleon matter from the inner core to the surface (dashed lines), $j_{max}$ can be larger than 0.7, mostly located in the range of $[0.7, 1.0]$, and the values are dependent sensitively on the star mass.
In addition, very similar conclusions about $j_{max}$ could be obtained for the hyperonic NS and the hybrid stars, as shown in Fig.~\ref{fig:3}(b) and Fig.~\ref{fig:3}(c) respectively.
Obviously, it is clear now that the crust structure around the star surface is an essential factor to determine the properties of the spin parameter of the compact stars, especially its maximum value. When the crust structure is considered, the traditional NS, hyperonic NS, and hybrid stars manifest almost the same behavior for the $j_{max}$, namely, the $j_{max}\sim 0.7$ for $M>0.5M_{\odot}$. Nevertheless, $j_{max}> 0.7$ if the crust structure is not included. From Fig.~\ref{fig:3}, one also find that the divergence between the curves with and without the crust disappears gradually as the star mass approaches to the maximum value, which is consistent with the discussion on Tab.~\ref{Tab1} and Tab.~\ref{Tab2}. Hence, a compact star source with smaller mass would be better discriminator to study the crust physics around the star surface. Moreover, it should be noticed that the strange quark stars are still the only one explanation to have $j_{max}>1$ which can be used to distinguish from the other compact stars. We also obtain the same and solid conclusions of $j_{max}$ by adopting various other crust EOSs, e.g., DH+HP EOS, which are not presented here.

\begin{figure}
\centering
\includegraphics[width=0.48\textwidth]{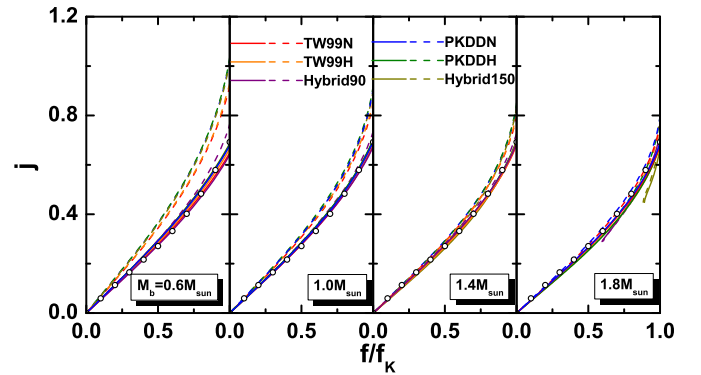}
\caption{The dimensionless spin parameter $j$ as a function of the scaled rotational frequency $f/f_K$ for the traditional neutron stars (PKDDN and TW99N), hyperonic neutron stars (PKDDH and TW99H) and hybrid stars (Hybrid90 and Hybrid150).
The baryonic mass of the stars is fixed at $M_b=0.6, 1.0, 1.4, 1.8M_{\odot}$, respectively.
The solid (dashed) line represents the results with (without) crust structure. The circles denote the reference values from the fitted formula $j=0.48(f/f_K)^3-0.42(f/f_K)^2+0.63(f/f_K)$. 
\label{fig:4}}
\end{figure}

Not only the maximum value, but also the results of the spin parameter $j$ at various rotational frequencies could be affected by the crust structure. The spin parameter $j$ as a function of the scaled rotational frequency $f/f_K$ for the traditional neutron stars (PKDDN and TW99N), hyperonic neutron stars (PKDDH and TW99H) and hybrid stars (Hybrid90 and Hybrid150) with (without) crust structure is displayed in Fig. \ref{fig:4}. The baryonic mass of the stars is fixed at $M_b=0.6, 1.0, 1.4, 1.8 M_{\odot}$, respectively. When crust is introduced (solid lines), it is surprising that the curves are almost unified for all the EOSs and mass sequences, namely independent on the structure of the interior of the compact stars and their mass. No matter which kind of composition is in the interior of the compact stars, such unified relationship maintains, which could be fitted approximately by the formula
\begin{equation}
j=0.48(f/f_K)^3-0.42(f/f_K)^2+0.63(f/f_K),
\end{equation}
as denoted with the circles in Fig. \ref{fig:4}. Such fitted universal formula could be used to deduce the third value when we have got two values of (spin parameter $j$, rotational frequency $f$ and keplerian frequency $f_K$).

When the crust structure is not included, it is clearly seen that the curves have a quite different tendency from the fitted formula. The deviation becomes larger for smaller mass sequence, same as the discussion on Tab.~\ref{Tab1} and Tab.~\ref{Tab2}. The largest deviation appears at the keplerian frequency for each fixed mass sequence. The results of the hybrid stars with $M_b = 1.8 M_{\odot}$ can not start from the origin because they belong to the supramassive sequence~\citep{5}. In short, this figure emphasizes the important role of the crust structure in the spin parameter once again, especially for the stars with smaller mass. Various other RMF EOSs are also utilized to study the crust effects on $j$, and it is verified that the results do not depend sensitively on the selections.


There are two points we want to emphasize from the above discussions:
1). The core contains up to 99\% of the mass of the compact stars while the crust contains less than 1\%~\citep{Lattimer04}. However, it is amazing to find that just this crust structure becomes a key factor to determine the properties of the spin parameter $j$, particularly its maximum value. Not the interior of the compact stars but the crust structure is one of the most important physical reason to the stability of $j_{max}$ for different kinds of compact stars.
2). All kinds of compact stars with crust structure have $j_{max}\sim 0.7$ for $M>0.5M_{\odot}$, while $0.7<j_{max}<1$ for the traditional NS, hyperonic NS and hybrid stars if the crust is not included. The strange quark stars with a bare quark-matter surface are the unique case to have $j_{max}>1$~\citep{8}.

Let us discuss briefly astrophysical implications of the results of the dimensionless spin parameter. \citet{Shibata03a, Shibata03b, Duez04, Piran05} suggested that the collapse of a uniformly rotating neutron star must have $j>1$ to form a massive disk around the final black hole, which further lead to the black-hole accretion model of gamma-ray burst.
Thus, \citet{8} suggested that the collapse of a rapidly rotating quark star could be a possible progenitor for the black-hole accretion model of gamma-ray burst. From our discussions, this conclusion is still reserved, as other kinds of compact stars hardly have $j>1$ for $M>0.5 M_{\odot}$ except the strange quark stars with a bare quark-matter surface.

Another issue is twin-peak quasi-periodic oscillations (kHz QPOs), which appears in low-mass X-ray binaries (LMXBs)~\citep{Gabriel12}. 
The relationship between $M$ and the spin parameter $j$ plays an important role in the understanding of the observed QPOs~\citep{Gabriel12, Gabriel12b, Gabriel12c}. In \citet{PASJ}, based on
a Resonantly Excited Disk-Oscillation Model, they suggested that the observed correlations in Cir X-1 could be well described by adopting $M = 1.5\sim2.0M_{\odot}$ and
$j\sim0.8$. Thus, \citet{8} suggested the central star in Cir X-1 could be a quark star if assuming Kato's model is correct, due to the calculated  uniformly rotating neutron stars with hadronic matter cannot have $j> 0.7$. However, in this work, we show that the $j_{max}$ of the traditional NS,  hyperonic NS and hybrid stars are also larger than 0.7 if the crust structure is not included. It seems that the possibility of the central star in Cir X-1 as a neutron star or a hybrid star can not be directly ruled out.

\section{Conclusions}

In conclusion, the dimensionless spin parameter $j$ of uniformly rotating compact stars has been investigated based on various
EOSs provided by the RMF theory and the MIT bag model. The results for different kinds of uniformly rotating compact stars, including the traditional NS, hyperonic NS and hybrid stars have been compared. It is shown that the crust structure is a key factor to determine the properties of the spin parameter of the compact stars. When the crust EOSs are considered, $j_{max}\sim 0.7$ for $M>0.5M_{\odot}$ is satisfied for three kinds of compact stars, no matter what the composition of the interior of the compact stars is. When the crust EOSs are not included, $j_{max}$ of the compact stars can be larger than 0.7 but less than about 1 for $M>0.5M_{\odot}$. Not the interior of the compact stars but the crust structure provide the physical origin to the stability of $j_{max}$ for different kinds of compact stars. The strange quark stars with a bare quark-matter surface are the unique one to have $j_{max}>1$. Thus, we can identify the strange quark stars based on the measured $j>1.0$, while measured $j\in(0.7,1.0)$ could not be treated as a strong evidence of the existence of a strange quark star any more. Furthermore, a universal formula $j=0.48(f/f_K)^3-0.42(f/f_K)^2+0.63(f/f_K)$ is suggested to calculate the spin parameter at any rotational frequency for all kinds of compact stars with crust structure and $M>0.5M_{\odot}$.


\begin{acknowledgements}
This work is partly supported by
the National Natural Science Foundation of China (Grant Nos. 11005069, 11175108, 11205075, 11375076),
the Specialized Research Fund for the Doctoral Program of Higher Education (Grant No. 20120211120002),
and the Shandong Natural Science Foundation (Grant No. ZR2010AQ005).
\end{acknowledgements}

\end{document}